\documentclass[letterpaper, 10 pt, conference]{ieeeconf}  

\IEEEoverridecommandlockouts                              

\usepackage[T1]{fontenc}
\usepackage{graphicx}

\title{\LARGE \bf
ASM: Audio Spectrogram Mixer
}

\author{Qingfeng Ji$^{1}$,  Jicun Zhang$^{2}$ and Yuxin Wang$^{3}$
\thanks{*This work was not supported by any organization}
\thanks{$^{1}$Qingfeng Ji is with School of Computer Science and Technology,
        Dalian University of Technology, Dalian, PRC
        {\tt\small 15640414255@mail.dlut.edu.cn}}%
\thanks{$^{2}$Jicun Zhang is with Neusoft Reach Automotive Technology(Dalian) Co.,Ltd.,
        Dalian, PRC
        {\tt\small zhangjicun89@163.com}}%
\thanks{$^{3}$Yuxin Wang is with School of Computer Science and Technology,
        Dalian University of Technology, Dalian, PRC
        {\tt\small wyx@dlut.edu.cn}}%
}

\begin{document}

\maketitle
\thispagestyle{empty}
\pagestyle{empty}

\begin{abstract}

Transformer structures have demonstrated outstanding skills in the deep learning space recently, significantly increasing the accuracy of models across a variety of domains. Researchers have started to question whether such a sophisticated network structure is actually necessary and whether equally outstanding results can be reached with reduced inference cost due to its complicated network topology and high inference cost. In order to prove the Mixer’s efficacy on three datasets—Speech Commands, UrbanSound8k, and CASIA Chinese Sentiment Corpus—this paper applies a more condensed version of the Mixer to an audio classification task and conducts comparative experiments with the Transformer-based Audio Spectrogram Transformer (AST) model. In addition, this paper conducts comparative experiments on the application of several activation functions in Mixer, namely GeLU, Mish, Swish and Acon-C. Furthermore, the use of various activation functions in Mixer, including GeLU, Mish, Swish, and Acon-C, is compared in this research through comparison experiments. Additionally, some AST model flaws are highlighted, and the model suggested in this study is improved as a result. In conclusion, a model called the Audio Spectrogram Mixer is suggested in this study, and the model’s future directions for improvement are examined.

\end{abstract}

\section{INTRODUCTION}

Encoder-Decoder structures and attention mechanisms havebeen extremely important in a number of artificial intelligence domains, including visual classification, temporal prediction, audio classification, etc., due to the popularity of Transformer structures in recent years. Due to their extensive parameters and intricate designs, transformer architectures have demonstrated excellent outcomes in data processing. Meanwhile, numerous models with intricate architectures have surfaced as a result of the huge rise in computing power and vast dataset collecting.

Researchers including Yuan Gong, Yu-An Chung, and
James Glass presented a model known as AST: Audio
Spectrogram Transformer at the Interspeech2021 conference(Gong, Chung, and Glass 2021). This model was born
out of the aforementioned situation. In the AST paradigm,
voice segments are converted into 128 The AST model converts speech snippets into 128-dimensional Mel spectral characteristics, and then uses a windowing operation
to produce the associated spectrograms. The spectrogram is
then divided into several segments by the AST model using
Transformer’s approach, which are then supplied into the encoder structure following Linear Projection and feature extraction. The categorization outcomes are then output by the
AST model using a linear layer. The model will be utilized
as a comparison target in this research since it exhibits good
performance on the three datasets.

In fact, the development of the Transformer structure has
greatly advanced deep learning. But given the high cost
of model inference, we begin to doubt the necessity of a
framework as intricate as Transformer. Google provided the
answer to this question in the paper titled ”MLP-Mixer:
An all-MLP Architecture for Vision” that was published in
NeurIPS(Tolstikhin et al. 2021). MLP-Mixer is the name of
the model architecture that they suggested. By combining
data in different positions (token-mixing) and in the same
data in different positions of the channel (channel-mixing),
the mixer is able to extract features easily. In short, feature
slices are extracted first in the column direction and then in
the row direction.

The Google team tested the effectiveness of MLP-Mixer
using various vision datasets and discovered that its relatively straightforward structure performs nearly as well as
various variations of Vision Transformer (ViT) while requiring less processing power. This shows that performance
comparable to that of complex structures can be obtained
even with simpler structures. The study ”MTS-Mixers: Multivariate Time Series Forecasting via Factorized Temporal
and Channel Mixing” released by the Huawei team in 2023
confirms the efficiency of the Mixer in the field of time series
forecasting(Li et al. 2023). The Huawei team emphasizes
that capturing temporal correlation does not always need attention. Their test findings on various real datasets demonstrate that MTS-Mixer outperforms existing Transformerbased algorithms in terms of efficiency.

We propose the question of whether the Mixer has the
potential to implement SOTA in the audio domain after validating the Mixer’s performance in the fields of vision and
timing prediction. We change the technical aspects of the
AST model that do not follow the objective laws and suggest a structure called Audio Spectrogram Mixer (ASM) by
using the Mixer

\section{ASM}

\subsection{Motivation}

According to validation on three datasets, AudioSet(Gemmeke et al. 2017), ESC50 (Piczak 2015),and Speech CommandsV2 (35 categories)(Warden 2018),the Transformer structure used by the AST model is capable of outperforming SOTA in the audio classification domain. This suggests that the AST model handles audio categorization tasks with greater performance. The authors of the research have developed an upgraded version of the AST model called ”SSAST: Self-Supervised Audio Spectrogram Transformer,” which enhances the performance of downstream audio tasks by utilizing masks and self-supervised
pre-training(Gong et al. 2022).

Based on the success of the AST model, we choose the
AST structure as a blueprint for our ASM model with the
structure shown in Figure 1 and compare the two on three
publicly available datasets. We aim to verify that the Mixer
has the potential to achieve SOTA in the audio classification
domain. By comparing with the AST model, we will further
explore the advantages and potential of the Mixer in the audio domain with the aim of bringing better performance to
audio classification tasks.

In addition to the above, we found one more aspect
of the AST model that needs improvement. The Transformer structure of the AST model is migrated from the ViT
model(Dosovitskiy et al. 2020) in the vision domain, which
deals with three-channel images, while the audio spectrogram has only single-channel features. To solve the problem
of channel number mismatch, the AST model directly sums
the weights of the three channels in the Linear Projection
layer to obtain the single channel weights. However, according to objective laws, we think this approach is not appropriate. Therefore, in this paper, we choose to use the formula
of RGB to grayscale map instead of this operation to deal
with the problem of channel number mismatch more accurately. With this alternative, we hope to further improve the
performance of the model and ensure consistency with the
objective law.

\subsection{Model Architecture}

\begin{figure}[t]
    \centering
    \includegraphics{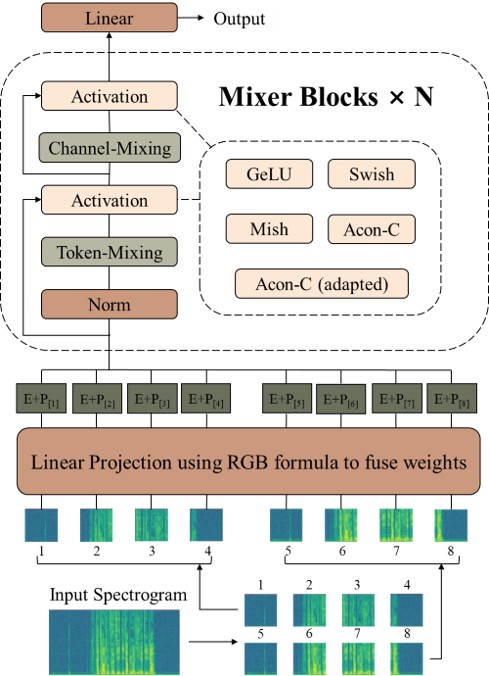}
    \caption{Audio Spectrogram Mixer}
    \label{fig:enter-label}
\end{figure}
Figure 1 shows a detailed illustration of the ASM architecture. First, the input audio is converted into 128-dimensional Mel-spectral features and windowed to obtain the corresponding spectrogram as the input to the ASM. Then, we divide the spectrogram into multiple patch sequences of size 16x16. These patch sequences are converted into 1D patch embeddings of size 768 by linear projection layer. next, the
embedded sequences are input into MLP-Mixer. In MLP-Mixer, we keep some settings similar to AST, such as keeping the embedding dimension at 768 and keeping 12 MLP-Mixer layers. In the MLP unit, we use GeLU(Hendrycks and Gimpel 2016), Mish(Misra 2019), Swish(Ramachandran,
Zoph, and Le 2017), and Acon-C(Ma et al. 2021) activation functions, which have non-zero gradients in the negative region and avoid the problem of dead neurons. In addition these functions are smoother at 0 than activation functions such as ReLU, and thus converge more easily during training.

Unlike the AST model, we modify the Mixer architecture
instead of choosing the original, unmodified Mixer architecture. Such modifications include adjustments to the shape of
the Mixer Blocks. We make this choice because there are
no pre-trained Mixer-like models available that are comparable to the number of DeiT(Touvron et al. 2021) parameters used by the AST model. Making comparisons in the
absence of substantial prior knowledge is as unfair as comparing the ability of high-potential newborns and junior high
school students to acquire knowledge in the same situation.
However, we retain a clear separation from the projection
layer at the input of the Mixer so that it can be replaced in
the future with a possible pre-trained Mixer-like model with
a large amount of prior knowledge.

\section{Experiments}

In AST, the authors provide an option to decide whether to
use a visual pre-training model (e.g., ImageNet equals True
or False), and the AST model uses a DeiT with a parametric
number of 87M and an image input size of 384x384 as its
visual pre-training model. Therefore, the RGB to grayscale
map formula is used in the Linear Projection layer only
when ImageNet=True.

To compare the capabilities of the Mixer and Transformer
structures, we choose to conduct comparison experiments in
the two cases where ImageNet is True and False. Among
them, when ImageNet is True, in order to avoid unfairness
due to DeiT’s Encoder structure with prior knowledge, we
reinitialize an Encoder unit with the same structure as DeiT,
but keep the weight parameter of DeiT before the Encoder
layer and use it as the baseline. the purpose of this is to ensure fairness and eliminate the influence of prior knowledge
on the experimental results.

It is also worth noting that the focus of the comparison experiments is only on verifying the capabilities of the Mixer
and the validity of the RGB to grayscale map formula, so we
don’t spend as much time fine-tuning and adapting the ASM
on AudioSet as AST does.

In our experiments, we use an ASM structure with a maximum number of parameters of 71M, which can save more
than 15\% of adjustable parameters and exhibit higher parameter efficiency compared to the AST model.

\subsection{Speech Commands Experiments} 
\subsubsection{Dataset Introduction and training details}

Speech Commands V2(Warden 2018) is an audio dataset containing 35
classifications, consisting of 84843 training samples, 9981
validation samples and 11005 test samples. The duration of
each sample is 1 second.

According to the settings in ”AST: Audio Spectrogram
Transformer”, we set the initial learning rate to 2.5e-4 and
reduce the learning rate of each epoch to 0.85 of the learning
rate of the previous epoch after the 5th epoch. for the two
cases of ImageNet being True and False, we respectively We
conducted experiments with 10 epochs and 30 epochs for the
two cases of ImageNet being True and False, respectively.
We select the best model by evaluating the metrics on the
validation dataset and report the evaluation metrics on the
test set. For the validation process of each model, we used
three different random seeds.

We choose the accuracy rate (ACC) and area under the
curve (AUC) as evaluation metrics.

\subsubsection{Speech Commands Results }

First, we conducted experimental validation for the case of not using the pre-trained
model, specifically, setting ImageNet to False and evaluating the effectiveness of replacing the Encoder with the
MLP-Mixer. The experimental results are detailed in Table
1, which lists the data for the validation set accuracy (v-acc),
the area under the validation set curve (v-auc), the test set accuracy (t-acc), and the area under the test set curve (t-auc).
Throughout the experiments in the paper, the metric names
and random seed labels are consistent to ensure comparable
experimental results.

Notably, after 30 epochs of training, we observe that
the ASM model performs significantly better than the AST
model (Baseline) in terms of ACC and AUC on the validation and test sets. This indicates that the replacement of
Encoder using the MLP-Mixer with the removal of the pretrained model can effectively improve the performance of
the model on this dataset.

\begin{table}[]
    \centering
    \caption{ImageNet=False, SCV2, AST and Mixer}
    \label{tab:1}
    \begin{tabular}{ccc}
         \hline
          & AST           & Mixer         \\ \cline{2-3} 
    v-acc & 0.8062±0.0088 & 0.9292±0.0084 \\
    v-auc & 0.9888±0.0007 & 0.9977±0.0006 \\
    t-acc & 0.7924±0.0093 & 0.9189±0.0106 \\
    t-auc & 0.9873±0.0009 & 0.9973±0.0005 \\ \hline
    \end{tabular}
\end{table}

Next, we verify the effectiveness of using the RGB to
grayscale map formula in Linear Projection under the condition of using a pre-trained model, i.e., ImageNet=True.
The experimental results are detailed in Table 2, where the
column and row names are the same as in the previous paper. We observe that after 10 epochs of training, the model
using the RGB-to-grayscale map formula performs consistently better than the AST model (Baseline) in terms of accuracy (ACC) and area under the curve (AUC) on both the
validation and test sets.

\begin{table}[]
    \centering
    \caption{ImageNet=True, SCV2, AST ,AST-RGB}
    \label{tab:2}
    \begin{tabular}{ccc}
         \hline
          & AST           & AST-RGB         \\ \cline{2-3} 
    v-acc & 0.8006±0.0215 & 0.8146±0.0101 \\
    v-auc & 0.9883±0.0020 & 0.9907±0.0011 \\
    t-acc & 0.7834±0.0301 & 0.7951±0.0167 \\
    t-auc & 0.9873±0.0019 & 0.9889±0.0015 \\ \hline
    \end{tabular}
\end{table}

At the end of this subsection, we further validate the effectiveness of using the MLP-Mixer to replace Encoder under the condition of using a pre-trained model (i.e., ImageNet=True). The experimental results are detailed in Table
3, where the column and row names have the same meaning as before. We observe that after 10 epochs of training,
the ASM model with the MLP-Mixer replacing Encoder performs significantly better than the AST model (Baseline) in
terms of accuracy (ACC) and area under the curve (AUC) on
the validation and test sets.

\begin{table}[]
    \centering
    \caption{ImageNet=True, SCV2, AST and Mixer}
    \label{tab:3}
    \begin{tabular}{ccc}
         \hline
          & AST           & Mixer         \\ \cline{2-3} 
    v-acc & 0.8006±0.0215 & 0.9470±0.0022 \\
    v-auc & 0.9883±0.0020 & 0.9987±0.0003 \\
    t-acc & 0.7834±0.0301 & 0.9401±0.0041 \\
    t-auc & 0.9873±0.0019 & 0.9983±0.0002 \\ \hline
    \end{tabular}
\end{table}

\subsection{UrbanSound8K Experiments}
\subsubsection{Dataset Introduction and training details}

UrbanSound8K(Jaiswal and Patel 2018) is a commonly used
audio classification dataset for research and experiments in
sound classification and ambient sound recognition. It consists of a series of audio samples from urban environments
and contains 8,732 short audio clips.

The following are some important information and features of the UrbanSound8K dataset:

Data source: Audio samples for the UrbanSound8K
dataset are collected from 10 different urban environments,
including streets, highways, parks, shopping malls, and more. These audio clips are collected by deploying sensors
in the city and recording environmental sounds.

Audio classification: Each audio clip in the dataset is classified into one of 10 categories, which include: air conditioner, car horn, children playing, dog bark, drilling, engine
idling, gunshot, jackhammer, siren, and street music.

Data format: Each audio clip is stored in WAV format
(lossless audio) and has the same sample rate (44.1 kHz).
The duration of the audio clips is between 1 and 4 seconds.

Dataset division: The UrbanSound8K dataset is divided
according to a training set and a test set. The training set
contains the majority of the 8,732 audio clips (with at least
400 samples per category), while the test set contains additional audio clips for evaluating model performance.

Metadata information: In addition to the audio samples
themselves, the dataset also contains metadata information
for each audio clip, such as file ID, audio category, sample
rate, duration, etc. This metadata information can be used to
train the model and perform evaluation.

The UrbanSound8K dataset provides a rich and diverse
sample of urban environmental sounds that can be used to
conduct research related to sound classification, environmental sound recognition, machine learning and deep learning. It has become one of the benchmark datasets for many
audio classification algorithms and models.

In this part of the experiments, we set the initial learning
rate to 1e-4 and after the 5th epoch, the learning rate of each
epoch decreases to 0.85 of the learning rate of the previous
epoch. we conducted two cases of experiments with ImageNet as True and False, and 10 epochs of training were performed in each case. We selected the best model by validating the evaluation metrics of the dataset and reported these
evaluation metrics on the test set. To increase the reliability of the experimental results, we performed three different
random seed validations for each model, as in the previous
section. We chose accuracy (ACC) as the primary evaluation metric and area under the curve (AUC) as a secondary
evaluation metric to assess the performance of the models.

\subsubsection{UrbanSound8K Results}

First, we verified the effectiveness of using MLP-Mixer to replace Encoder without using
the pre-trained model (i.e., ImageNet = False). The results
of our experiments are shown in Table 4, where the column
and row names have the same meaning as before.

By comparing the experimental results, we can observe
that after 10 epochs of training, the ASM model using MLP-Mixer to replace Encoder performs significantly better than
the AST model (Baseline) on the validation set.

\begin{table}[]
    \centering
    \caption{ImageNet=False, US8K, AST and Mixer}
    \label{tab:4}
    \begin{tabular}{ccc}
         \hline
          & AST           & Mixer         \\ \cline{2-3} 
    v-acc & 0.7628±0.0089 & 0.8852±0.0029 \\
    v-auc & 0.9676±0.0015 & 0.9906±0.0012 \\
    t-acc & 0.7510±0.0113 & 0.8951±0.0069 \\
    t-auc & 0.9662±0.0018 & 0.9903±0.0019 \\ \hline
    \end{tabular}
\end{table}

Next, we verify the effectiveness of using the RGB-to-grayscale map formula in Linear Projection and the effectiveness of using the MLP-Mixer to replace Encoder in the
case of using a pre-trained model (i.e., ImageNet = True).
The experimental results are shown in Table 5, where the
meanings of column and row names are the same as before.

After 10 epochs of training, we observe that the model
using the RGB-to-grayscale map formula performs stably
better than the AST model (Baseline) on the validation and
test sets.

\begin{table}[]
    \centering
    \caption{ImageNet=True, US8K, AST ,AST-RGB}
    \label{tab:5}
    \begin{tabular}{ccc}
         \hline
          & AST           & AST-RGB         \\ \cline{2-3} 
    v-acc & 0.8806±0.0053 & 0.8932±0.0043 \\
    v-auc & 0.9892±0.0015 & 0.9908±0.0010 \\
    t-acc & 0.8898±0.0024 & 0.8951±0.0074 \\
    t-auc & 0.9896±0.0014 & 0.9905±0.0013 \\ \hline
    \end{tabular}
\end{table}

In addition, we can also find that the ASM model using
MLP-Mixer to replace Encoder performs significantly better
than the AST model (Baseline) on the validation and test
sets.

\begin{table}[]
    \centering
    \caption{ImageNet=True, US8K, AST and Mixer}
    \label{tab:6}
    \begin{tabular}{ccc}
         \hline
          & AST           & Mixer         \\ \cline{2-3} 
    v-acc & 0.8806±0.0053 & 0.9230±0.0142 \\
    v-auc & 0.9892±0.0015 & 0.9944±0.0020 \\
    t-acc & 0.8898±0.0024 & 0.9176±0.0127 \\
    t-auc & 0.9896±0.0014 & 0.9944±0.0009 \\ \hline
    \end{tabular}
\end{table}

\subsection{CASIA Chinese Sentiment Corpus Experiments}
\subsubsection{Dataset Introduction and training details}

The CASIA
Chinese Sentiment Corpus(Ke et al. 2018) is recorded by the
Institute of Automation, Chinese Academy of Sciences. The
corpus covers speech material recorded by four professional
pronouncers, including six emotions: angry, happy, fear, sad,
surprised and neutral. A total of 9600 speech samples with
different pronunciations are included.

Of these, 300 speech samples are recorded based on the
same text, i.e., the same text is given different emotional
states for reading aloud. These samples can be used to compare and analyze the acoustic characteristics and rhythmic
performance in different emotional states.

In addition, there are 100 speech samples are recorded
based on different texts. These texts are literally clear in their
emotional attribution, enabling the pronouncer to express the
corresponding emotional state more accurately.

The CASIA Chinese Sentiment Corpus is designed to
provide researchers with rich phonetic data to explore differences in the acoustic and rhythmic aspects of emotional
expression. The data in this corpus are recorded from professional speakers and possess high quality and accuracy,
which can provide strong support for research in the fields
of emotion recognition and speech emotion processing.

In this section, we refer to the setup of the Speech Commands V2 dataset experiments. The initial learning rate is set
to 2.5e-4, and after the 5th epoch, the learning rate of each
epoch decreases to 0.85 of the learning rate of the previous epoch. we conduct experiments with 25 epochs in both the True and False cases of ImageNet. The best model is
selected by validating the evaluation metrics of the dataset,
and we report these evaluation metrics on the test set. To increase the reliability of the experimental results, we perform
three different random seed validations for each model. We
choose the accuracy rate (ACC) of the model.

\subsubsection{CASIA Chinese Sentiment Corpus Results}

First, we
verify the effectiveness of using the MLP-Mixer to replace
the Encoder without using the pre-trained model (i.e., ImageNet = False). The results of our experiments are shown
in Table 7, where the column and row names have the same
meaning as before.

By comparing the experimental results, we can observe
that after 25 epochs of training, the ASM model using MLP-Mixer to replace Encoder performs significantly better than
the AST model (Baseline) in terms of accuracy (ACC) and
area under the curve (AUC) on the validation and test sets.

\begin{table}[]
    \centering
    \caption{ImageNet=False, CASIA, AST and Mixer}
    \label{tab:7}
    \begin{tabular}{ccc}
         \hline
          & AST           & Mixer         \\ \cline{2-3} 
    v-acc & 0.7833±0.0194 & 0.9025±0.0089 \\
    v-auc & 0.9783±0.0033 & 0.9948±0.0012 \\
    t-acc & 0.7760±0.0112 & 0.8997±0.0107 \\
    t-auc & 0.9780±0.0024 & 0.9943±0.0008 \\ \hline
    \end{tabular}
\end{table}

Next, we verify the validity of using the RGB to grayscale
map formula in Linear Projection in the case of using a pretrained model (i.e., ImageNet = True). The experimental results are shown in Table 8, where the column and row names
have the same meaning as before.

After 25 epochs of training, we observe that the model using the RGB-to-grayscale map formula performs relatively
consistently better than the AST model (Baseline) in terms
of accuracy (ACC) and area under the curve (AUC) on both
the validation and test sets.

\begin{table}[]
    \centering
    \caption{ImageNet=True, CASIA, AST ,AST-RGB}
    \label{tab:8}
    \begin{tabular}{ccc}
         \hline
          & AST           & AST-RGB         \\ \cline{2-3} 
    v-acc & 0.7932±0.0121 & 0.7984±0.0044 \\
    v-auc & 0.9821±0.0021 & 0.9834±0.0016 \\
    t-acc & 0.7997±0.0099 & 0.8102±0.0064 \\
    t-auc & 0.9826±0.0012 & 0.9839±0.0007 \\ \hline
    \end{tabular}
\end{table}

At the end of this subsection, we validate the effectiveness
of using the MLP-Mixer to replace the Encoder for the case
where a pre-trained model is used (i.e., ImageNet = True).
The experimental results are detailed in Table 9, where the
meaning of the column and row names in the table is the
same as in the previous section.

After 25 epochs of training, we observe that the ASM
model using MLP-Mixer to replace Encoder performs significantly better than the AST model (Baseline) in terms of
accuracy (ACC) and area under the curve (AUC) on both the
validation and test sets.

\begin{table}[]
    \centering
    \caption{ImageNet=True, CASIA, AST and Mixer}
    \label{tab:9}
    \begin{tabular}{ccc}
         \hline
          & AST           & Mixer         \\ \cline{2-3} 
    v-acc & 0.7932±0.0121 & 0.9247±0.0056 \\
    v-auc & 0.9821±0.0021 & 0.9962±0.0010 \\
    t-acc & 0.7997±0.0099 & 0.9176±0.0088 \\
    t-auc & 0.9826±0.0012 & 0.9945±0.0012 \\ \hline
    \end{tabular}
\end{table}

\subsection{Comparison of Activation Functions in Mixer}
\subsubsection{Dataset and Experimental setup}

We use the CASIA Chinese Sentiment Corpus as the experimental dataset for this
part, and use the Gelu activation function used in ”MLP-Mixer” as the baseline, and use the Mish, Swish and Acon-C activation functions for comparison. In addition, we make
appropriate adjustments to the function characteristics of
Acon-C to ensure that the p1 parameter is initialized near
1 and the p2 parameter is initialized near 0 as much as possible. Since the implementation of Acon-C requires the use
of batch size as a parameter, considering the number of data
samples in the CASIA Chinese sentiment corpus, we made
appropriate adjustments to ensure that the impact of batch
size on the experiment is eliminated in order to make the experiment more complete. We still use 3 random seeds in our
experiments, which is the same as described in the previous
section.

The initial learning rate is set to 2.5e-4 and after the 5th
epoch, the learning rate of each epoch decreases to 0.85 of
the learning rate of the previous epoch. we conduct experiments with ImageNet = False for 25 epochs. The best model
was selected using evaluation metrics from the validation
set, and we report these evaluation metrics on the test set.
We choose accuracy (ACC) as the main evaluation metric
and area under the curve (AUC) as a secondary evaluation
metric.

\begin{figure}[t]
    \centering
    \includegraphics{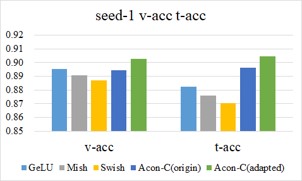}
    \caption{Seed-1, ACC}
    \label{fig:enter-label}
\end{figure}

\subsubsection{Experimental Results}

After 25 epochs of training, the results of our experiments on random seed No. 1 are shown
in Table 10, Figure 2 and Figure 3. From them, it can be
observed that the model with the use of the adjusted Acon-C activation function performs best in all four evaluation-metrics. However, the Swish activation function performs the
worst in all four evaluation-metrics.

\begin{table*}[]
    \centering
    \caption{ImageNet=True, CASIA, Seed-1}
    \label{tab:10}
    \begin{tabular}{clllcc}
         \hline
     Seed-1 & GeLU   & Mish   & Swish  & Acon-C & Acon-C(adapted) \\ \hline
    v-acc  & 0.8954 & 0.8907 & 0.8870 & 0.8944 & 0.9028          \\
    v-auc  & 0.9922 & 0.9927 & 0.9922 & 0.9926 & 0.9952          \\
    t-acc  & 0.8824 & 0.8759 & 0.8704 & 0.8963 & 0.9046          \\
    t-auc  & 0.9915 & 0.9914 & 0.9911 & 0.9916 & 0.9934          \\ \hline
    \end{tabular}
\end{table*}

\begin{figure}[h]
    \centering
    \includegraphics{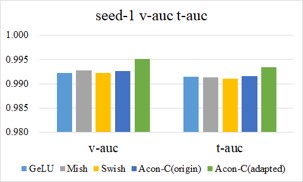}
    \caption{Seed-1, AUC}
    \label{fig:enter-label}
\end{figure}

\begin{figure}[h]
    \centering
    \includegraphics{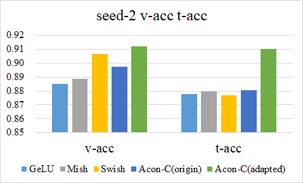}
    \caption{Seed-2, ACC}
    \label{fig:enter-label}
\end{figure}

\begin{figure}[h]
    \centering
    \includegraphics{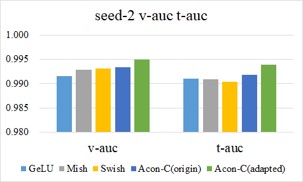}
    \caption{Seed-2, AUC}
    \label{fig:enter-label}
\end{figure}

Similarly, the results of our experiments with random seed
No. 2 are shown in Table 11, Figure 4, and Figure 5, and it is
found that the model using the adjusted Acon-C activation
function still performs best on the four metrics.

\begin{table*}[]
    \centering
    \caption{ImageNet=True, CASIA, Seed-2}
    \label{tab:11}
    \begin{tabular}{clllcc}
         \hline
     Seed-1 & GeLU   & Mish   & Swish  & Acon-C & Acon-C(adapted) \\ \hline
    v-acc  & 0.8852 & 0.8889 & 0.9065 & 0.8972 & 0.9120          \\
    v-auc  & 0.9915 & 0.9928 & 0.9931 & 0.9933 & 0.9949          \\
    t-acc  & 0.8778 & 0.8796 & 0.8769 & 0.8806 & 0.9102          \\
    t-auc  & 0.9910 & 0.9908 & 0.9904 & 0.9919 & 0.9939          \\ \hline
    \end{tabular}
\end{table*}

In the experiments with random seed No. 3, the results
of the evaluation metrics for the validation and test sets are
shown in Table 12, Figure 6, and Figure 7, and we again find
that the model using the adapted Acon-C activation function
performs best on all four metrics.

\begin{table*}[]
    \centering
    \caption{ImageNet=True, CASIA, Seed-3}
    \label{tab:12}
    \begin{tabular}{clllcc}
         \hline
     Seed-1 & GeLU   & Mish   & Swish  & Acon-C & Acon-C(adapted) \\ \hline
    v-acc  & 0.8815 & 0.8963 & 0.9009 & 0.8833 & 0.9139          \\
    v-auc  & 0.9926 & 0.9930 & 0.9935 & 0.9915 & 0.9948          \\
    t-acc  & 0.8759 & 0.8843 & 0.8824 & 0.8787 & 0.9019          \\
    t-auc  & 0.9900 & 0.9906 & 0.9909 & 0.9904 & 0.9947          \\ \hline
    \end{tabular}
\end{table*}

\begin{figure}[h]
    \centering
    \includegraphics{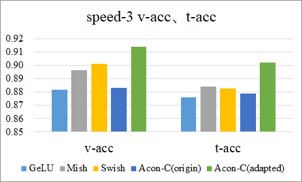}
    \caption{Seed-3, ACC}
    \label{fig:enter-label}
\end{figure}

\begin{figure}[h]
    \centering
    \includegraphics{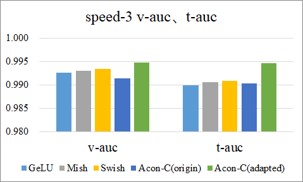}
    \caption{Seed-3, AUC}
    \label{fig:enter-label}
\end{figure}

\section{Conclusion and Follow-up}
\subsection{Conclusion}

In recent years, the Transformer has become an indispensable and critical component in various fields. However, in
our study, we find that it is not necessary to adopt such a
complex network structure. In the field of vision and temporal prediction, there are already successful cases that demonstrate the superiority of the Mixer, which reduces the inference cost (according to our experiments, saving more than
20\% of the training time compared to AST) while having
an accuracy that is as good as or even better than that of
the Transformer. The Mixer does not yet have a pre-trained
model comparable to the volume of DeiT, so we cannot verify whether Mixer can maintain its excellent performance
with a large amount of prior knowledge. However, based on the work in this paper, we have good reasons to believe that
Mixer has the potential to challenge Transformer in the field
of audio classification, and researchers should pay more attention and devote more resources to the research of MLP
and Mixer.

Also in this study, we validate the use of the RGB-to-grayscale map formulation instead of simple weighted fusion. The experimental data show that the optimized model
performs stably better than the original model, although the
improvement in accuracy is only between 1\% and 3\%. We
should recognize that the accuracy of the model is mainly
influenced by factors such as model structure and data volume. However, this optimization work still reminds us that
changes and optimizations should be made with more reference to objective laws and knowledge that has been practically validated in other fields. Such an approach can provide
us with more targeted directions for improvement, which
can lead to better model performance.

\subsection{Follow-up}

We plan to apply the existing Mixer pre-trained network
to ASM to explore its effectiveness in audio classification
tasks. Compared with Transformer, Mixer does not seem to
require trainable location information, so we consider verifying whether removing the CLS flag in AST will have an
impact on the accuracy. In addition, we plan to further optimize the structure of the Mixer unit to find a more suitable
Mixer for audio classification tasks.

Besides, we are also inspired by the SSAST method and
plan to try to train the ASM model using a self-supervised
training approach to solve the problem of mismatch between the migrated visual model and the audio spectrogram.
This will help improve the generalization performance of the
model and further enhance the accuracy of audio classification.

\addtolength{\textheight}{-12cm}   









\end{document}